**The impact of motor and non-motor symptoms fluctuations on health-related quality of life in people with functional motor disorder**

Martin Jirásek[1,4], Tomáš Sieger, PhD[1,2], Gabriela Chaloupková, PhD[1], Lucia Nováková[1], Petr Sojka, PhD[1], Mark J Edwards, Prof., PhD[3], Tereza Serranová, PhD[1]

1. Department of Neurology and Center of Clinical Neuroscience, First Faculty of Medicine, Charles University and General University Hospital in Prague, Prague/Czech Republic. Electronic address: tereza.serranova@vfn.cz

2. Department of Cybernetics, Faculty of Electrical Engineering, Czech Technical University, Prague/Czech Republic

3. King´s College London, Institute of Psychiatry, Psychology & Neuroscience, Department of Basic & Clinical Neuroscience, London/United Kingdom

4. Department of Rehabilitation and Sports Medicine, Second Faculty of Medicine, Charles University and University Hospital Motol, Prague/Czech Republic

**Corresponding Author:**

Tereza Serranova, MD, PhD

First Faculty of Medicine, Charles University and General University Hospital in Prague, Department of Neurology and Centre of Clinical Neuroscience

Kateřinská 30, 12000 Prague 2, Czech Republic

phone: +420 22496 5539, e-mail: tereza.serranova@vfn.cz

**Abstract**

**Objective:** To assess the effect of overall, between- and within-day subjectively rated fluctuations in motor and non-motor symptoms in people with functional motor disorder (FMD) on the health-related quality of life (HRQoL).

**Background:** FMD is a complex condition characterized by fluctuating motor and non-motor symptoms that may negatively impact HRQoL.

**Methods:** Seventy-seven patients (54 females, mean age 45.4 ± 10.4 years) with a clinically established diagnosis of FMD, including weakness, completed symptom diaries, rating the severity of motor and non-motor symptoms (i.e., pain, fatigue, mood, cognitive difficulties) on a 10-point numerical scale three times daily for seven consecutive days. HRQoL was assessed using the SF-36 questionnaire. For the analysis, fluctuation magnitude was defined in terms of the variability in self-reported symptom scores.

**Results:** The mental component of SF-36 was jointly predicted by the overall severity scores ($t(74) = -3.61$, $P < 0.001$) and overall general fluctuations ($t(74) = -2.98$, $P = 0.004$). The physical SF-36 was found to be related only to the overall symptom severity scores ($t(74) = -7.09$, $P < 0.001$), but not to the overall fluctuations. The assessment of the impact of different components showed that the mental component of SF-36 was significantly influenced by the combined effect of average fatigue ($t(73) = -3.86$, $P < 0.001$), between-day cognitive symptoms fluctuations ($t(73) = -3.22$, $P = 0.002$), and within-day mood fluctuations ($t(73) = -2.48$, $P = 0.015$).

**Conclusions:** This study demonstrated the impact of self-reported symptom fluctuations across multiple motor and non-motor domains on mental but not physical HRQoL in FMD and highlighted the importance of assessing and managing fluctuations in clinical practice.

**Introduction**

Functional movement disorders (FMD) are highly prevalent conditions[1,2] which are associated with significant disability [3], poor health-related quality of life (HRQoL) [3], and high healthcare costs [4]. FMD is clinically characterised by the inconsistency of motor symptoms, manifesting as variability between tasks and/or within one task [5]. Demonstration of the variability of motor signs, for example, distractibility of functional tremor or transient resolution of unilateral leg weakness with contralateral hip flexion (Hoover's sign), is a key part of the positive diagnosis of FMD [6]. Moreover, historical features of symptom waxing and waning over the long term and fluctuations during the day are found more frequently in the FMD group than in other neurological conditions [7]. Besides motor symptoms, most people with FMD also have multiple non-motor symptoms, such as anxiety, depression, cognitive symptoms, fatigue, and pain, which are generally rated by patients as having a greater impact on quality of life than the severity of motor symptoms [8–10].

Fluctuations in both motor and non-motor symptom severity can significantly impact daily activities and overall well-being. Unpredictable changes in symptom severity may restrict individuals with FND in planning and managing their daily routines, leading to increased stress and a negative impact on HRQoL [11–16]. The importance of symptom fluctuations has already been documented in various neurological diseases and diseases with functional symptoms [13,17–19]. In FMD, symptoms may fluctuate over different periods of time, with within-day and between-day changes in symptom severity potentially having distinct effects on HRQoL. Despite their reported importance, these phenomena have not been studied in FMD. A better understanding of how fluctuations in symptom severity and their influence on HRQoL is essential for improving FMD assessment and management. Additionally, it could help in the important task of developing FND-specific outcome measures [20,21].

We hypothesised that a higher sum of fluctuations in the motor and key non-motor symptom severity would be associated with a lower HRQoL in both the physical and mental symptoms in individuals with FMD. Additionally, we hypothesised that fluctuations in symptom severity, when analysed in conjunction with the severity of motor and non-motor symptoms, would be an independent predictor of HRQoL outcomes in individuals with FMD.

To assess the impact of fluctuations, including within-day and between-day symptom fluctuation, on HRQoL, we collected diaries with self-reported symptom severity in motor and key non-motor symptoms and HRQoL measures in a group of people with FMD. The impact of fluctuations in symptom severity was assessed in addition to the average symptom severity. For exploratory purposes, we also examined the relative importance of fluctuations in symptom severity alongside individual symptom severity to better understand their distinct and combined effects on predicting HRQoL in individuals with FMD.

**Methods**

**Participants**

Ninety individuals with FMD meeting the inclusion criteria were recruited from the specialised neurological center of the General University Hospital in Prague. The study was approved by the local ethics committee (approval no. 37/19) and all participants gave their written consent to take part in the study. The inclusion criteria stipulated that participants must be 18 years of age or older and have a clinically established diagnosis of FMD, according to Gupta and Lang criteria made at specialised outpatient service for FMD at the Neurology Department. The diagnosis of FMD was established based on positive signs of internal inconsistency in abnormal movements or weakness observed within a task or between different tasks, using general and phenotype-specific tests.[6,22].

Exclusion criteria were defined as: age <18 years, inability to complete questionnaires because of language difficulties, severe learning disabilities or cognitive impairment, psychosis, or neurological comorbidity associated with sensorimotor impairment.

**Clinical assessment**

Clinical assessments included neurological examination with phenotyping of all motor symptoms according to the predominant and additional types of abnormal movements (i.e., weakness, tremor, dystonia, gait abnormalities, myoclonus, and speech and swallowing problems), and recording of concomitant medication. Examiner-based assessment of motor symptoms severity was performed using the Simplified Functional Movement Disorders Rating Scale (S-FMDRS) [23]. In this scale, the presence or absence of abnormal movement at each of seven body regions (face and tongue, head and neck, left upper limb and shoulder girdle, right upper limb and shoulder girdle, trunk and abdomen, left lower limb, right lower limb) were rated according to symptom severity and duration (maximum score: 54).

**Subjective evaluation of symptom fluctuations**

All subjects were given colour-printed paper questionnaires with instructions on how to complete them over the next consecutive days. They were asked to rate subjective symptom severity across all individual symptoms on a 0 to 10 scale, where 0 denoted the absence of symptoms, and 10 signified the most severe manifestation of symptoms. Patients recorded their ratings for motor and non-motor symptoms, including pain, fatigue, mood, and cognitive difficulties, three times daily (morning, afternoon, and evening) on the provided paper diaries. These non-motor symptoms were selected based on previous evidence of their impact on HRQol [8–10]. The colour-coded system, utilising shades of blue (the darker the colour, the more severe the symptoms), visually complemented this numerical scale, facilitating a comprehensive assessment of symptom severity over the 10-day observation period.

**Outcome measures**

On day 10, after completing the diary, patients filled out the 36-item Short Form Health Survey questionnaire (SF-36) to assess HRQoL [24]. The SF-36 questionnaire consists of eight scales yielding two outcome measures: physical component score of SF-36 (hereafter referred to as physical SF-36) and mental component score of the SF-36 (hereafter referred to as mental SF-36). The physical SF-36 includes four scales of physical functioning (10 items), role-physical (4 items), bodily pain (2 items), and general health (5 items). The mental SF-36 is composed of vitality (4 items), social functioning (2 items), role-emotional (3 items), and mental health (5 items). A final item, termed self-reported health transition, is answered by the client but is not included in the scoring process. Each summary measure is directly transformed into a 0-100 scale, assuming that each question carries equal weight. The lower the SF-36 scores, the more disability in each component.

**Symptom Severity and Fluctuations Analysis**

For the statistical analysis, we used average scores of individual and overall symptom severity ratings computed from 5 individual symptoms, including movement difficulties, mood (depression and anxiety), cognition (memory and concentration problems), pain, and fatigue.

**The average score of an individual symptom severity** was defined as the average value of that symptom (i.e., the average of 21 ratings collected over seven days and three times a day).

**The overall score of symptom severity** was defined as the average diary value of all symptoms (i.e. the average of 105 values collected over seven days and three times a day across five symptom domains).

**We defined fluctuation magnitude in individual and overall symptom severity** in terms of the variability of self-reported symptom scores on the 10-point Likert scale using a statistical method enabling to differentiate between two distinct types of fluctuations. To measure how much symptoms vary from one day to another (between-day fluctuations) and over the course of a single day from morning to evening (within-day fluctuations), we fitted a two-way ANOVA model, which looks at the effects of both the day and the time of day (Fig. 1).

**Between-day fluctuations of an individual symptom** were defined as the square root of the mean square associated *with the effect of the day* in a two-way ANOVA model with the effects of the day and time of the day.

**Within-day fluctuations of an individual symptom** were defined as the square root of the mean square associated *with the effect of the day (morning-afternoon-evening)* in a two-way ANOVA model with the effects of the day and time of day. Within-day fluctuations represented systematic average daily progression of symptom severity. The slope of a linear fit to the three values of the effect of the time of day was taken as a measure of average improvement or worsening of symptoms over a day.

**Residual fluctuations** were defined as the difference between the self-reported symptom scores and the sum of between-day and within-day effects. Residual fluctuations represented the random (irregular) part of the symptom scores.

**General fluctuations of an individual symptom** were defined as the sum of between-day, within-day, and residual fluctuation.

**Overall fluctuations (general, between-day, or within-day, respectively)** were defined as the average of the respective fluctuations over all individual symptoms.

**Statistics**

The relation between symptoms (scores and their fluctuations) and QoL was assessed using linear models. When the number of predictors was greater than three (i.e. when scores and fluctuations of five individual symptoms entered the model), we used a lasso model, i.e., a penalised maximum likelihood linear model, to prevent overfitting. The penalisation parameter was estimated using 10-fold cross-validation, taking the largest value for which its associated cross-validation error was within one standard error of the value, yielding minimal cross-validation error. Nested linear models were compared using analysis of variance assessing the reduction in their residual sum of squares.

**Results**

All 90 patients with motor FMD who met inclusion criteria underwent a full clinical assessment and agreed to complete the questionnaires. However, in 13 patients, data from questionnaires were missing (all of them completed less than seven full days in symptom severity diary). All subjects with missing data were excluded from the analysis.

Seventy-seven patients (54 females, mean age 45.4 (standard deviation, SD = 10.4, range 18.1 - 62.4 ) years; mean disease duration: 5.7 (SD = 5.2) years) completed diaries for at least seven days and completed SF-36 questionnaire on day 10.

Demographic and clinical characteristics are presented in Table 1.

*Table 1.*

| FMD patients | |
|---|---|
| Subjects N | 77 |
| Age (years) | 45.4 (±10.4) |
| Females N | 54 |
| FMD duration (years) | 5.7 (±5.2) |
| S-FMDRS | 14.3 (±6.8) |
| Motor phenotype present as dominant/present total | N (%)/N (%) |
|     Tremor | 15 (19.5)/41 (53.2) |
|     Gait disorder | 31 (40.3)/64 (83.1) |
|     Weakness | 21 (27.3)/58 (75.3) |
|     Dystonia | 7 (9.1)/23 (29.9) |
|     Myoclonus | 2 (2.6)/6 (7.8) |
|     Speech and swallowing disorder | 1 (1.3)/10 (13.0) |
|     Parkinsonism | 0 (0.0)/1 (1.3) |

Note: Abbreviations: FMD, functional movement disorder; SD, standard deviation; S-FMDRS, Simplified Functional Movement Disorder Rating Scale

In patients in whom symptoms changed over the course of the day, the day effects of all symptoms mostly worsened over the day (in 72% of patients for pain, in 82% for fatigue, in 65% for mood, in 76% for cognitive symptoms, and in 69% for motor symptoms).

First, we analysed how **overall** scores and overall general fluctuations could jointly explain mental and physical SF-36. We found that the mental SF-36 was related to both the overall scores and overall general fluctuations (Fig. 2A,B). Notably, the overall scores and overall general fluctuations, respectively, were found to combine into a single strong predictor of the mental component of SF-36 in a linear model with adjusted $R^2$ of 0.245 ($t(74)=-3.61$, $P<0.001$, and $t(74)=-2.98$, $P=0.004$, respectively). The mental SF-36 decreased, on average, by 2.5 points for an increase of 1 point of the overall symptom severity score and by 3.9 points for an increase of 1 point of the overall general fluctuations.

The physical SF-36 was found to be related only to the overall symptom severity scores ($t(74)=-7.09$, $P<0.001$, Fig. 3A), but not to the overall general fluctuations ($t(74)=-0.17$, $P=0.87$) (Fig. 3B)

Additionally, we wanted to learn which components of overall scores and overall fluctuations were related to SF-36. First, we analysed how average scores, between-day, within-day, and residual fluctuations of individual symptoms combined to explain the mental component of SF-36 using a lasso model. The model identified average fatigue, between-day and residual fluctuations of cognitive symptoms, and within-day fluctuations of mood as relevant predictors. As between-day and residual fluctuations of cognitive symptoms were highly correlated and negatively affected each other in the model, we omitted the residual fluctuations of cognitive symptoms from the model without affecting the model quality (F(1,73)=2.18, P=0.14). We learned that the combined effect of average fatigue, between-day fluctuations of cognitive symptoms, and within-day fluctuations of mood significantly influenced the mental SF-36 (t(73)=-3.86, P<0.001, t(73)=-3.22, P=0.002, and t(73)=-2.48, P=0.015, resp., Fig. 2C, D). The mental SF-36 decreased, on average, by 2.0 points for an increase of 1 point of fatigue severity, by 4.7 points for a unit increase of between-day fluctuations of cognitive symptoms, and, simultaneously, by 3.6 points for a unit increase in within-day fluctuations of mood.

Looking at individual symptoms, the physical component of SF-36 could best be explained by the average motor symptoms score (t(75)=-8.12, P<0.001). The physical component of SF-36 decreased, on average, by 5.2 points for a unit increase in the motor symptom severity score (Fig. 3C).

**Discussion**

This study, using self-reported diary-based assessments of motor and key non-motor symptoms, provided evidence for different subtypes of symptom fluctuations and demonstrated their impact on HRQoL in individuals with FMD. We found that symptom severity fluctuations significantly impacted mental HRQoL, whereas physical HRQoL was influenced only by overall symptom severity. The combined effect of overall symptom severity and general fluctuations strongly predicted mental HRQoL. Between-day fluctuations had a more pronounced effect on mental HRQoL. The combined influence of average fatigue, between-day fluctuations in cognitive symptoms, and within-day fluctuations in mood significantly affected mental QoL. The physical component of the SF-36 was primarily associated with average motor symptoms. These findings have important implications for symptom assessment and management strategies.

In line with previous research in FMD, we found a relationship between motor and non-motor symptom severity and HRQoL [8–10]. Here, we demonstrated that mental HRQoL is also influenced by overall symptom fluctuations. The overall (average) symptom severity and fluctuations are complementary; of two patients with the same average symptom severity, the one with the higher fluctuations will have a lower mental quality of life; on the other hand, of two patients with the same fluctuations magnitude, the one with the higher average symptom severity will have a lower mental and physical quality of life. Mental QoL decreases from about 65 to 30 points with increasing overall symptom severity (Fig 2ABC), while physical QoL decreases from about 60 to 0 (Fig 3AC).

Most of the quantitative studies in chronic fatigue syndrome and pain conditions have primarily focused on how fluctuations in various non-motor symptoms interact and affect physical activity levels [11,12,15], while only a few studies addressed the impact of non-motor symptom fluctuations on HRQoL. Fluctuations have also been assessed in other neurological disorders, such as multiple sclerosis or Parkinson's disease. In multiple sclerosis, intraindividual increases and decreases in symptom severity, such as depression, anxiety, and fatigue, recorded longitudinally over longer periods of time (annually for three years) correlated with changes in HRQoL [16]. In Parkinson's disease, the fluctuations in motor and non-motor symptoms are the hallmark of advanced disease and are targeted by numerous pharmacological strategies and advanced therapies such as deep brain stimulation. Interestingly, the impact of fluctuations in Parkinson's disease has been studied mostly in terms of the percentage/duration of the disabling OFF time periods but not in terms of fluctuation magnitude [17]. In FND, fluctuations are likely driven by different mechanisms, reflecting the unique pathophysiology of the disorder, and can be targeted by behavioural interventions.

An important strength of this study is the use of a statistical method to define and differentiate distinct types of symptom fluctuations in a larger group of FMD patients, allowing us to assess their impact on HRQoL. We found that between-day fluctuations had a larger impact on HRQoL. However, both between-day and within-day fluctuations contributed to mental health. While this is an important finding, it is difficult to discuss our results in the context of previous literature. Previous studies largely varied not only in populations being studied data collection methods (time periods, recording tools) but also in defining symptom variability/fluctuations as an outcome measure. Numerous studies have not presented a clear definition of fluctuations [18,19,28,29]. Two studies in chronic fatigue syndrome, including within-day fluctuations, used averaged symptom severity, and the variations (fluctuations) were measured by calculating the standard deviation of these scores, a method similar to ours [11,12].

We did not find a relationship between fluctuations and self-reported physical health reflecting general mobility, among other items. An objective assessment of physical activity using accelerometers along with subjective measures could disentangle the complex relationship between physical activity, non-motor symptom severity, and HRQoL in FMD. In a study on chronic fatigue syndrome, physical activity measured using accelerometry correlated with the severity of fatigue, pain, and mood [11]. However, it has been repeatedly demonstrated that subjective and objective measures rarely correlate in FMD [30]. For example, a prior study on various tremor types, including functional tremor, found a discrepancy between tremor duration reported in diaries and actigraphic recordings, with patients generally overestimating their tremor. This mismatch was more pronounced in those with functional tremor [31].

The impact of symptom fluctuations over time on various aspects of mental HRQoL, including psychological well-being, social interactions, work capability, and challenge-making commitments, could result from the unpredictability of symptoms' severity in FMD. In a qualitative study of individuals with chronic pain, the timing of the onset of pain flares, the severity of pain flares, and fluctuations in pain severity were prioritised as being key features of a pain forecast, and making plans were prioritised as being a key benefit [32]. Many of the symptoms assessed in our study, such as fatigue, cognition, and affective state physiologically, also fluctuate in healthy individuals. Given the lack of recordings from healthy controls, we can only speculate about the physiological and well-tolerated size of fluctuations in different physical and mental symptoms, including pain, fatigue, mood, and cognitive symptoms. The evidence for a higher fluctuation in functional symptoms severity compared to other disorders is very limited. One study reported higher fluctuations in fibromyalgia compared to osteoarthritis and rheumatoid arthritis [15].

Our findings highlight the need for clinical assessment and management of between-day and within-day fluctuations in FMD patients on a routine basis. Keeping track of symptom severity over the long term could offer a more accurate understanding of the "boom and bust" cycles often seen in FNDs [33,34]. A complete assessment should consider both motor and non-motor symptom fluctuations, including fatigue, cognitive, and mood changes, to create a personalised multimodal treatment plan. Stabilising symptom fluctuations can be achieved through education on the "boom and bust" cycles and using techniques such as activity planning, pacing, and graded activity. While the pacing and graded activity could help improve fatigue [35], further studies are needed to assess whether cognitive rehabilitation and behavioural or pharmacological approaches address cognitive and mood fluctuations in this population [36–38]. However, when gathering information about symptom fluctuations over longer time periods in retrospective clinical assessment, it is important to consider that a patient's report can be influenced by their current state [39]. To gain a more realistic understanding of the patient's long-term health, it might be necessary to collect data continuously using diaries or wearables that can help to track symptoms.

Our findings also have implications for the development of FND-specific outcome measures. Since both motor and non-motor symptom fluctuations have an impact on HRQoL, it is worth considering incorporating measures of within-day and between-day fluctuations into the development of patient-reported outcome measures assessing illness severity [20,32].

This study has limitations, including a lack of a control group with healthy subjects or a different clinical population with multiple motor and non-motor symptoms, such as individuals with multiple sclerosis, which are also known to have symptom fluctuations. Additionally, with the use of technology, further studies allowing assessment of various objective and subjective measures will be beneficial to achieve a better understanding of fluctuations across different symptom domains. The non-inclusion of other symptoms, such as bladder or bowel dysfunction, dizziness, or psychological symptoms like dissociation, now increasingly recognized as important and frequent in this population, also represents a limitation of the study and warrants further investigation. Future studies should explore the impact of fluctuations over extended periods (days, weeks, or months) using wearable technology, which could offer a more feasible and comprehensive approach to capturing these dynamics in real-world settings. Additionally, further research should investigate whether distinct fluctuation patterns are associated with specific clinical profiles in larger cohorts, as such studies could provide valuable insights into potential subgroup distinctions within this population.

In conclusion, this study found an association between symptom fluctuations and HRQoL in people with FMD. Self-reported symptom fluctuations across motor and non-motor domains had significant impact on mental but not physical HRQoL. Mental HRQoL was strongly predicted by both overall severity and symptom fluctuations, with between-day fluctuations having a greater impact. These findings have important clinical implications for clinical assessment and treatment of FMD. Additionally, these results provide important input for developing patient-reported outcome measures for FND.

**Conflict of Interest:** The authors have no conflict of interest to report.

**Funding:** This work was supported by the Czech Ministry of Health Project AZV (NU20-04-0332 and NU24-04-00456), General University Hospital in Prague (MH CZ-DRO-VFN64165); The project National Institute for Neurological Research (Programme EXCELES, ID Project No. LX22NPO5107) - Funded by the European Union – Next Generation EU); and ERDF-Project Brain Dynamics, No. CZ.02.01.01/00/22_008/0004643.

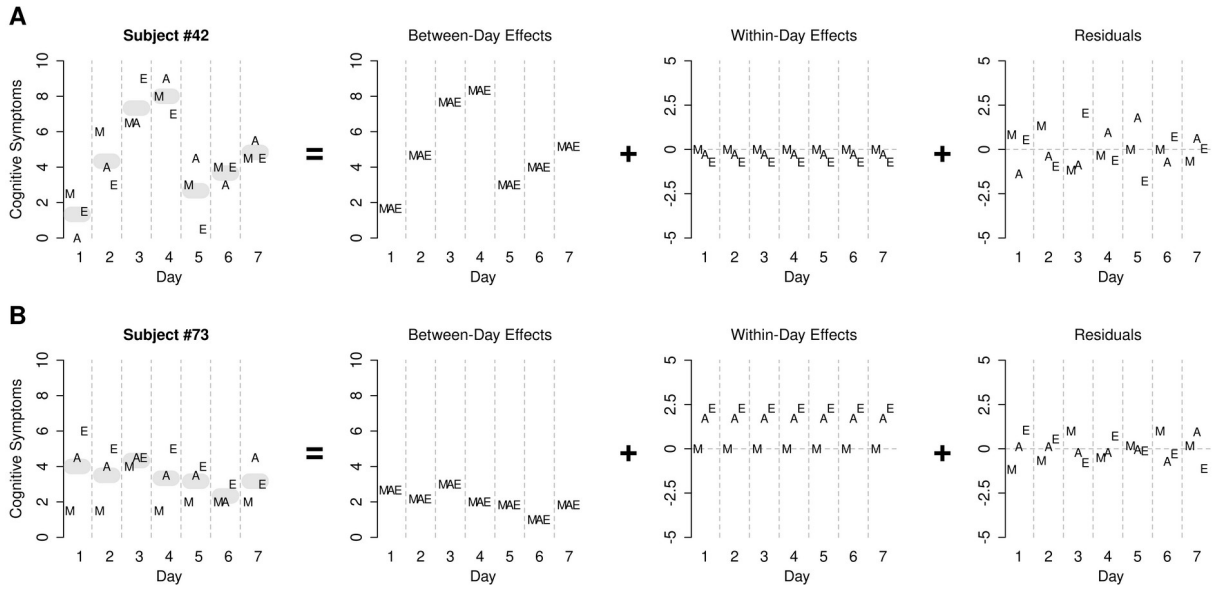

**Fig. 1** An example decomposition of cognitive symptom severity into between-day effects, within-day effects, and residuals. Morning, afternoon, and evening symptom severities are depicted with M, A, and E, respectively. The grey ovals denote day averages.

(A) In subject #42, the severity of cognitive symptoms varied mostly over days, as between-day effects (2nd column) were larger compared to within-day effects (3rd column). (B) In subject #73, the severity of cognitive symptoms varied mostly within days. Note that in patient #42, the cognitive symptoms severity decreased, on average, during the day, while in patient #73, the cognitive symptoms worsened over the course of the day.

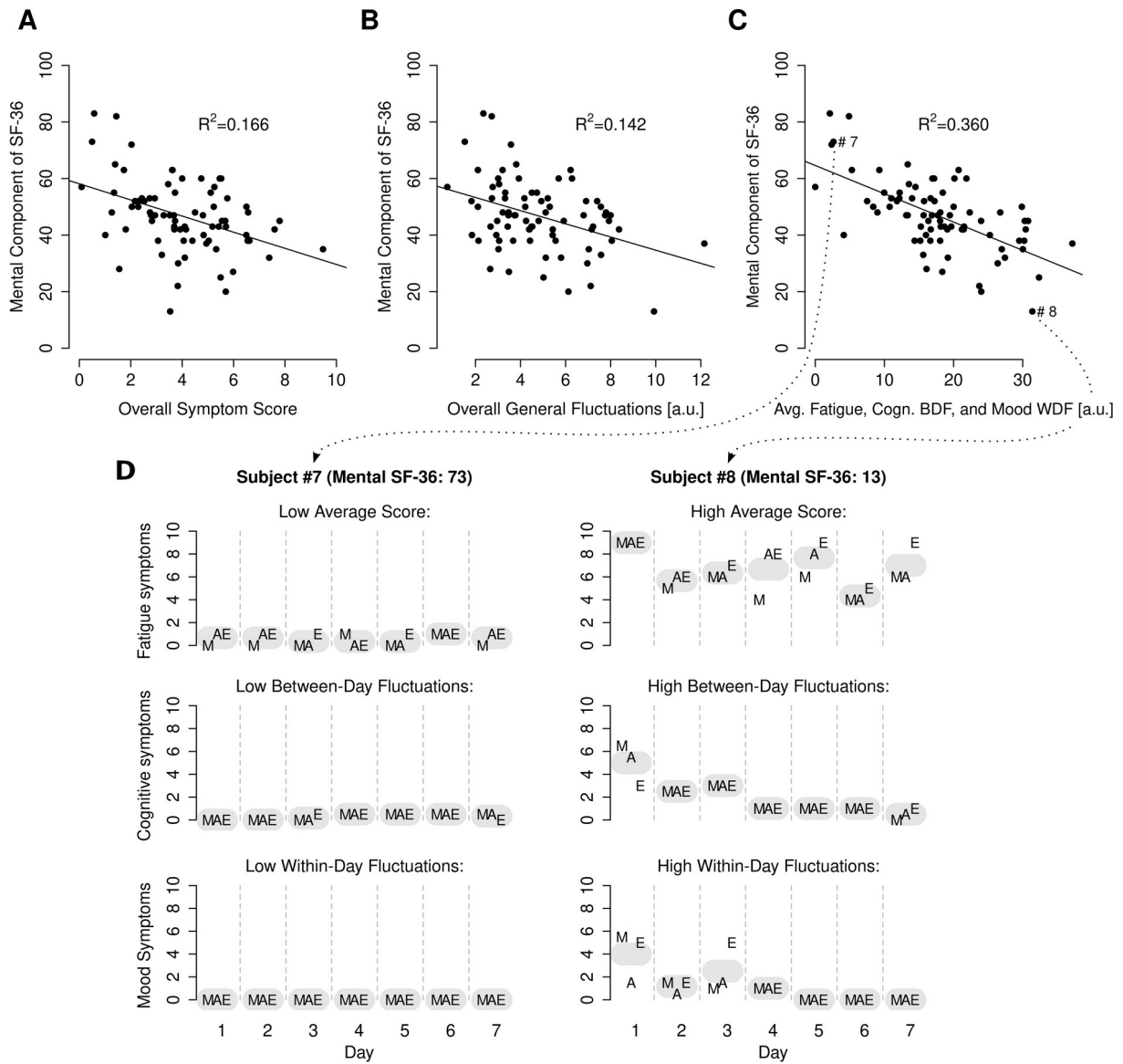

**Fig 2.** The mental component of SF-36 is related to the combined effect of symptom severity scores and their fluctuations.

(A) The mental component of SF-36 decreased by 2.9 points with a unit worsening of the overall symptom severity scores. (B) The mental SF-36 component decreased by 4.7 points, with a unit worsening in the overall general fluctuations of the symptom severity scores. (C) The mental SF-36 component could best be explained using a combined predictor consisting of the mean fatigue severity score, between-day fluctuations of the cognitive symptoms, and within-day fluctuations of the mood symptoms.

(D) Illustration of mean fatigue severity score, between-day fluctuations of the cognitive symptoms, and within-day fluctuations of the mood symptoms affecting the mental component of SF-36. Note that subject #7 with mental SF-36 of 73, has relatively low fatigue, low between-day fluctuations of cognitive symptoms, and low within-day fluctuations of mood compared to subject #8, with mental SF-36 of 13.

Morning, afternoon, and evening symptom severity is depicted with M, A, and E, respectively. The grey ovals denote day averages. $R^2$ are adjusted coefficients of determination expressing the predictive strength of individual models. BDF = between-day fluctuations, WDF = within-day fluctuations.

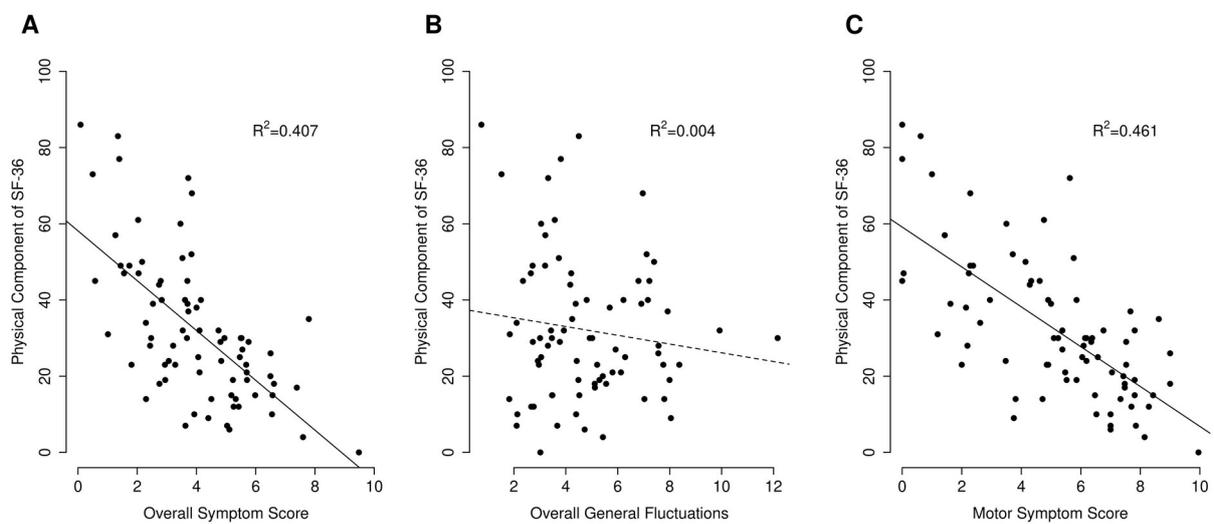

**Fig 3.** The physical component of SF-36 is related only to symptom severity scores.

(A) The physical component of SF-36 decreases by 6.5 points with a unit worsening of the overall symptom severity scores. (B) The physical component of SF-36 was not related to overall general fluctuations of the symptom severity scores. (C) The physical component of SF-36 could best be related to the severity of movement symptoms.